# Rethinking Competition as a Non-Beneficial Mechanism in Economic Systems

**Abstract**

Persistent economic competition is often justified as a mechanism of innovation, efficiency, and welfare maximization. Yet empirical evidence across disciplines reveals that competition systematically generates fragility, inequality, and ecological degradation, emergent outcomes not of isolated failures but of underlying systemic dynamics. This work reconceptualizes economic ecosystems as real complex adaptive systems, structurally isomorphic with biological and social ecosystems. Integrating complexity science, evolutionary biology, ecology, and economic and business theory, we classify economic interactions according to their systemic effects and propose a theoretical model of ecosystemic equilibrium based on the predominance of beneficial versus non-beneficial relationships. Recognizing economies as ecologically embedded and structurally interdependent systems provides a novel framework for analyzing systemic resilience, reframing competition as a non-beneficial mechanism.

**Marcelo S. Tedesco** $^\delta$     **Gonzalo Marquez**$^\phi$

---

$^\delta$**Affiliation:**
Buenos Aires Institute of Technology (ITBA), Buenos Aires, Argentina.
Massachusetts Institute of Technology (MIT), MIT LIFT Lab, Cambridge, MA, United States.
Corresponding Author: tedesco@itba.edu.ar; tedesco@mit.edu

$^\phi$ **Affiliation:**
National University of La Plata (UNLP), Faculty of Natural Sciences and Museum; La Plata, B1900; Argentina.

# Introduction

The organization of human economic behaviour has long been guided by the assumption that competition drives progress, innovation, and collective welfare. Since the publication of *The Wealth of Nations* (*1*), mainstream economic theory has positioned competition as the central mechanism aligning individual self-interest with societal benefit, framing markets as self-regulating systems capable of achieving equilibrium through decentralized action (*2–7*).

Yet in practice, sustained competitive dynamics across modern economies increasingly produce emergent patterns of systemic fragility, inequality, and environmental depletion. These outcomes are not incidental anomalies but consistent expressions of how interdependent agents behave within complex adaptive systems—where non-linearity, feedback loops, and structural dependencies amplify destabilizing interactions over time (*8–15*). This Perspective proposes a transdisciplinary framework grounded in complexity science, evolutionary biology, and ecological systems theory to re-examine the systemic role of competition, reconceptualizing it as a structurally non-beneficial interaction that undermines long-term resilience within nested economic, social, and biological ecosystems.

However, this competitive paradigm has proven inadequate for balancing economic development with social equity and environmental sustainability. Empirical research over the past decades confirms that modern capitalism, driven by competitive dynamics, fails to operate within the planet's biophysical boundaries while simultaneously exacerbating social and economic asymmetries (*16-18*).

Even studies that defend competition recognize the need for compensatory mechanisms to prevent overexploitation and inequality (*19-21*). Piketty (*18*) explicitly shows that competitive capitalism—left unchecked—tends toward structural concentration of wealth and political power. A recent study shows that "superstar firms" emerge not solely through innovation, but through feedback loops that reinforce inequality and limit diffusion of benefits—hallmarks of systemic disequilibrium (*20*).

Although some defenders of market competition argue that externalities and inequality are correctable anomalies, this paper takes a different view. We argue that these are not anomalies but emergent properties—the natural and recurrent result of structurally non-beneficial interactions in economic ecosystems. Pollution, financial crises, monopolies, and social polarization are not deviations from a healthy system, but outcomes of an architecture based on persistent competition, parasitism, and predation. In this sense, what mainstream economics refers to as "market failures" or "externalities" are better understood as signals of systemic disequilibrium emerging from the interactional structure of complex adaptive systems (*22-27*).

This critique aligns with a growing body of literature advocating for a shift from reductionist economic models to frameworks grounded in complexity science (*28-30*). These approaches conceptualize the economy not as a closed, equilibrium-seeking mechanism, but as a real complex adaptive system, embedded in social and ecological networks, and defined by emergent and interdependent dynamics (*28-33*).

Despite these advances, economic and business theory has historically imported concepts from biology and systems science only selectively and metaphorically—failing to adopt their full epistemological implications (*24, 34-38*). While the term "ecosystem" has gained traction in business and policy discourse since Moore's (*39*) introduction and subsequent elaborations by Cohen (*40*), Autio & Thomas (*41*), and other scholars (*42, 43*), it is most often treated as a metaphor or borrowed label, devoid of systemic rigor (*44*).

This paradox is rarely acknowledged: foundational biological concepts—such as competition, mutualism, and cooperation—emerged from sociological inquiry rather than from empirical observation alone (*45-47*). While biology matured into an empirical science grounded in systems thinking, economic theory remained anchored in doctrinal models derived from its dominant schools of thought (*34-37*).

However, ecosystems—biological, social, and economic—share near-identical systemic properties, including emergence, feedback, non-linearity, and dynamic equilibrium (*8*, 31, *48-52*). These shared characteristics support a more precise conclusion: economic ecosystems are not analogous to biological ones—they are structurally isomorphic.

This paper embraces a transdisciplinary approach, integrating insights from complexity science, ecology, evolutionary biology, and philosophy of science, to reconceptualize economic ecosystems as structurally real, not metaphorical.

**A transdisciplinary review of ecosystems and complexity in economics**

This study integrates insights from four disciplinary domains: evolutionary biology and ecology, economic and business sciences, complexity science, and philosophy of science. Biology and ecology provide empirical and theoretical foundations for interdependence and interaction types; economics and business capture institutional and agent-level dynamics. Complexity science explains systemic behaviors such as feedback and emergence, while philosophy of science supports integration across paradigms and ensures conceptual coherence.

To establish a robust theoretical foundation, we reviewed 168 publications drawn primarily from Science, Nature, Web of Science, JSTOR, SpringerLink, and ScienceDirect. The corpus includes 75 references from economics and business sciences, 46 from biological and ecological sciences, 37 from complexity science—with particular emphasis on applications in economics and biology—and 10 from philosophy of science.

Selection prioritized widely cited, conceptually innovative, and recent works, with particular emphasis on publications from the last three years (2022–2025) to meet current advances across disciplines. Additional recent studies from 2019 to 2021 were incorporated to capture the latest conceptual developments. Landmark works from 2001 to 2018 and foundational contributions prior to 2000 were further included to support theoretical integration, epistemological grounding, and cross-disciplinary clarification.

A methodological breakdown of the reviewed works is presented in Table 1.

Table 1. Methodological Composition of the Reviewed Literature

| *Disciplinary Domain* | *Total Works* | *Foundational Works* | *Last Works 2022-2025* | *Recent Works 2019-2021* | *Landmark Works 2001-2018* | Selection Criteria |
|---|---|---|---|---|---|---|
| *Biology and Ecology* | 46 | 5 | 12 | 6 | 23 | Influence in the Field, Theoretical or Empirical Support, Citations. |
| *Economic and Business Science* | 75 | 17 | 15 | 18 | 25 | Influence in the Field, Theoretical or Empirical Support, Citations. |
| Complexity Sciences | 37 | 7 | 7 | 10 | 13 | Influence in the Field, Theoretical or Empirical Support, Citations. |
| *Philosophy of Science* | 10 | 5 | 2 | 1 | 2 | Influence in the Field, Theoretical or Empirical Support, Citations. |

**Source:** Own elaboration

A fewer in number, contributions from the philosophy of science play a critical role by challenging the internal assumptions of dominant economic paradigms and supporting epistemological integration. Their inclusion reflects the premise that economic ecosystems are not just empirical objects, but theoretical constructs requiring cross-disciplinary coherence.

Our review applied critical, comparative, and integrative methods to identify both theoretical advances and gaps in how ecosystems are conceptualized across biology, economics, and systems theory. The findings underscore that integrating ecosystem theory into economics requires moving beyond metaphor and adopting a complexity-based systems perspective.

These insights provide the conceptual foundation for the next section, where we formalize the structural properties of complex systems and apply them directly to economic ecosystems.

*Complex systems theory as a framework for economic analysis*

Complex systems theory emerged to explain phenomena that defy reductionist analysis, emphasizing how interdependent components produce emergent behaviors not attributable to individual elements *(53–56)*. General Systems Theory, introduced by Bertalanffy *(57)*, laid the foundation by showing that biological and social complexity arises from the organization of interacting subsystems *(58, 59)*.

Across fields—from chaos theory in physics to network dynamics in biology—complexity science has developed a coherent framework for understanding how feedback, adaptation, and structural organization shape system behavior *(60–62)*. In ecology, these models explain population dynamics, trophic interactions, and resilience, highlighting how small perturbations can trigger large-scale transformations through feedback amplification *(63–66)*.

In economics, Arthur *(67)* showed how positive feedback can entrench dominance by specific technologies or firms, even when initial advantages are marginal. Later studies confirmed that such mechanisms contribute to market concentration, wage polarization, and reduced diffusion of innovation (20, 68). Similar dynamics affect inter-firm connectivity, supply chain fragility, and financial contagion, underscoring the systemic relevance of network structures *(15, 69)*.

Recent applications of complexity theory in developmental economics *(70)* and innovation ecosystems *(71)* show that decentralized agency and diversity enhance adaptive capacity. These findings align with ecological research linking diversity and interaction quality to systemic resilience *(72, 73)*.

Social sciences have also incorporated complexity theory. The "small-world" network model *(74)* showed how clustered yet globally connected social structures facilitate rapid diffusion of ideas and power consolidation—insights now central to political science, sociology, and innovation studies *(75)*.

Meadows *(31)* describes complex systems by three core features: elements, relationships, and function or purpose. Cities, forests, and economies all embody this structure *(8, 76–79)*. Multiple studies confirm that biological, social, and economic systems share near-exact structural properties—non-linearity, emergence, feedback, and interdependence—qualifying them as structurally isomorphic *(48–51)*.

In contrast, neoclassical and mainstream economic models—such as general equilibrium (80) and DSGE frameworks *(81)*—assume linearity, rational expectations, and representative agents. These approaches refine micro-level assumptions, but largely overlook emergent systemic behavior *(82–85)*.

While some orthodox economists argue that significant progress has been made in responding to earlier criticisms (86), and that these models still offer value in various applied settings *(87, 88)*, emerging works contest this view—portraying them as fundamentally irredeemable due to their failure to reflect the complex, adaptive nature of economic systems *(89–91)*.

We argue that a systems-based perspective—grounded in structural interdependence, feedback loops, and adaptation—is necessary for analyzing economic ecosystems. These systems operate through diverse interaction types, including mutualistic, cooperative, parasitic, and competitive, each exerting distinct effects on systemic equilibrium.

Understanding these dynamics requires a transdisciplinary lens—one that integrates methods and knowledge across fields to construct frameworks capable of addressing complex problems *(92, 93)*. Given their shared systemic architecture with biological and social systems, economic ecosystems demand this kind of conceptual synthesis.

## Ecological mechanisms as structural foundations

The concept of "ecosystem" entered the business and economic literature through Moore *(39)*, who used it to describe the co-evolution of firms in dynamic environments, invoking biological terminology to emphasize interdependence and adaptive change. Although initially metaphorical, the notion was rapidly adopted across studies of entrepreneurship and innovation systems (*40- 43*). Despite its increasing presence, most applications remain loosely defined, often framed in terms of "biological inspiration," "ecological metaphors," or models "derived from biology" (*94–97*). Even studies employing the language of complexity frequently stop short of grounding the concept in formal systems theory (*43, 98–100*).

This ambiguity has created a conceptual bottleneck. Reviews of the most-cited publications confirm that "ecosystem" is overwhelmingly treated as adapted from biology, not as a rigorously defined analytical construct (*44*). This lack of epistemological clarity has hindered comparative research, theoretical coherence, and the development of robust models for understanding interdependence and systemic behavior (*101, 102*).

Clarifying the concept requires moving beyond metaphor. The prefix "eco" itself derives not from biology but from the Greek οἶκος (*oikos*), meaning "house" or "dwelling," a root shared by *economy* and *ecology* (*103*). This common origin suggests that the application of "ecosystem" to economics is not merely metaphorical but potentially structurally valid—if analytically grounded.

Biological ecosystems offer a rich foundation for such grounding. Defined as systems of living organisms interacting with their physical environment, they are widely recognized as archetypes of complex adaptive systems (*48, 104, 105*). These systems exhibit emergent behavior, non-linearity, and feedback loops—properties also observed in economic networks (*31, 33, 106–107*). Their ability to maintain dynamic stability, or ecosystemic equilibrium, underpins their resilience and adaptability (*49, 73, 108*).

Ecosystemic equilibrium—also referred to in biology as *ecological homeostasis*—describes the dynamic stability that emerges from continuous interactions among organisms and their environment. This property allows ecosystems to persist despite internal perturbations or external shocks (*104, 109–111*). It reflects a system's capacity to absorb disturbances while maintaining its functional integrity over time.

The ability of an ecosystem—or any complex system—to shift between equilibrium states in response to such disturbances defines its resilience (*112, 113*). This resilience is reinforced by *structural robustness*, which emerges from strong and diverse interactions among agents. Such configurations enhance the system's long-term stability and its capacity to sustain life over time (*72, 73, 114–117*).

The nature of interactions among agents—whether mutually beneficial or harmful—is a primary driver of ecosystemic equilibrium. In ecological theory, such interactions are systematically categorized by their long-term systemic effects:

Table 2. Typology of Ecosystem Interactions and Their Long-Term Effects

| Type of Interaction and Long-Term Effect | Type of Interaction | Interaction Sign (Effect-Based) | Description of the Effect on the Participants |
|---|---|---|---|
| **Beneficial** | Mutualism | +/+ | All participating organisms benefit from the interaction, and none are harmed. It occurs between different species and involves a degree of interdependence. |
| | Cooperation | +/+ | All participating organisms benefit from the interaction, and none are harmed. It can occur within or between species and does not imply interdependence. |
| | Commensalism | +/0 | One or more organisms benefit from the interaction without negatively affecting any other. |
| | Facilitation | +/+/0 | One or more organisms modify the environment in ways that benefit others. |
| **Non-Beneficial** | Predation | +/- | One or more organisms benefit from the interaction while the others are harmed. |
| | Parasitism | +/- | One or more organisms benefit at the expense of others, which are harmed by the interaction. |
| | Amensalism | 0/- | One or more organisms are unaffected by the interaction, while others are harmed. |
| | Competition | -/- | All participants are negatively affected as they compete for limited resources. |

**Source:** Adapted by the authors from ecological interaction mechanisms (118–122).

The abundance and quality of beneficial relationships among diverse participants strengthen ecosystems' ability to absorb abrupt changes and maintain equilibrium (*73, 123, 124*). Mechanisms such as cooperation and mutualism are structurally central to both stability and adaptability (*26, 120, 125*).

Empirical studies confirm that resilient ecosystems are dominated by beneficial interactions. Mutualism, cooperation, and facilitation frequently account for over 70% of all observed relationships (*126, 127*), enhancing adaptability and buffering against external shocks (*123, 124*). In contrast, ecosystems where competition, predation, or parasitism prevail tend to exhibit reduced diversity and increased instability (121, *128–130*).

Although innovation theorists like Schumpeter (*2*) praised competition for its role in creative destruction, ecological evidence suggests that intense and exclusionary competition degrades resilience (121, *131–133*). It is diversity and the quality of interactions—not competition itself—that underpin equilibrium. These insights map directly onto economic systems, which also display path dependence, non-linear propagation, and feedback amplification (*19, 76, 134*).

Economic crises—including the Great Depression and the 2008 collapse—did not arise from isolated failures but from self-reinforcing feedback among agents and institutions (*14, 135*). As with ecological regime shifts, these collapses occurred when resilience thresholds were breached (*136*). In both ecosystems and economies, emergent phenomena stem from agent interdependence and structural topology—not from individual optimization alone (*32, 137*). Recognizing this, complexity economists have begun to model economies as adaptive systems (*8, 19, 28, 79, 134, 138*).

Ecological resilience theory offers corrective insight. Just as diversity reinforces robustness in biological systems (*115, 121*), distributed agency and network complexity increase the resilience of economic systems (*70*). Conversely, competitive exclusion and resource concentration—exemplified by the rise of "superstar firms"—replicate ecological patterns of systemic fragility (*20*). This is not a metaphorical resemblance but a manifestation of shared systemic properties: adaptive complexity, feedback amplification, and structural vulnerability.

A systems-based framework—grounded in the structural classification of interactions—offers a more coherent way to analyze the dynamics of real economic ecosystems. Beneficial relationships foster systemic equilibrium, while non-beneficial ones, particularly sustained competition, generate fragility. In modern capitalism, the prevalence of the latter undermines resilience and exacerbates both inequality and ecological degradation. Recognizing economies as real complex adaptive systems—subject to the same structural constraints as biological and social ecosystems—provides a stronger foundation for theoretical analysis and policy response in the face of mounting systemic risk.

## Grounding economic ecosystems in complex systems theory

Building on the previously mentioned conceptual ambiguity, economic ecosystems—whether entrepreneurial, business-oriented, or innovation-driven—have often been described as "inspired by" or "metaphorically derived from" biology (33, 44, 139). While such early framing contributed to their initial diffusion, it simultaneously limited theoretical coherence and empirical applicability by failing to establish formal systemic foundations (101, 102, 140).

However, a growing body of research points to structural similarities that go beyond metaphor. Studies across economics and complexity science identify core features of economic systems—such as emergent properties, adaptive feedback, interdependence, and non-linearity—not as superficial analogies, but as indicators of shared systemic architecture (*43, 71, 98, 100,*). Economic agents—firms, individuals, institutions—operate within environments shaped by natural and social constraints, exchanging resources in ways that generate systemic dynamics (8, 28, 30, 32, 33, 49, 67, 77, 141)

The systemic embeddedness of economic activity has long been highlighted by scholars who argue that economic dynamics cannot be fully understood without recognizing their integration into social systems. In this view, every economic system operates as a sub-system of the social system, which itself is embedded within the broader biological system (*23, 36, 141–144*).

Building on this research, we contend that economic ecosystems should not be seen as analogies, but as real complex adaptive systems—structurally isomorphic with biological and social ecosystems.

We propose the following definition:

> An economic ecosystem is a community of interdependent agents who interact dynamically with one another and with their environment, within a region bounded by social and natural constraints, where resources are continuously exchanged with the function and/or purpose of generating economic value in various forms.

This definition builds upon prior conceptualizations (*33, 43, 145*) by integrating key components from systems theory—agents, environment, resources, function or purpose, and interactions (*31*). It also incorporates human agency, feedback loops, and biophysical flows, aligning economic systems with ecological ecosystems and reinforcing their status as non-metaphorical, structurally grounded systems.

This alignment is not only theoretical. It is empirically observable through the material metabolism of economic systems. Economy-wide material flow accounting offers a dynamic lens to examine how resources, stocks, and waste circulate within the socioeconomic domain (*141, 146*). Ultimately, all inputs sustaining economic activity originate in—and return to—the biosphere, as demonstrated by ecological research on system functioning and service provision (*141, 147, 148*).

Economic behavior and policy decisions also feed back into the social system, shaping well-being and the distribution of resources. Research on austerity and development models illustrates how economic structures both transform and are conditioned by the social fabric (14, *149*), highlighting the system's recursive and adaptive nature.

Framing economic systems as panarchically nested within social and biological ecosystems reveals their multilevel interdependencies (*49, 142, 141, 150, 151*). As illustrated in Fig. 1, these nested structures interact through cross-scale feedback loops and structural couplings. The economy depends not only on continuous biophysical throughput (*144–154*), but also co-produces social configurations that shape systemic resilience and influence long-term sustainability (*149*).

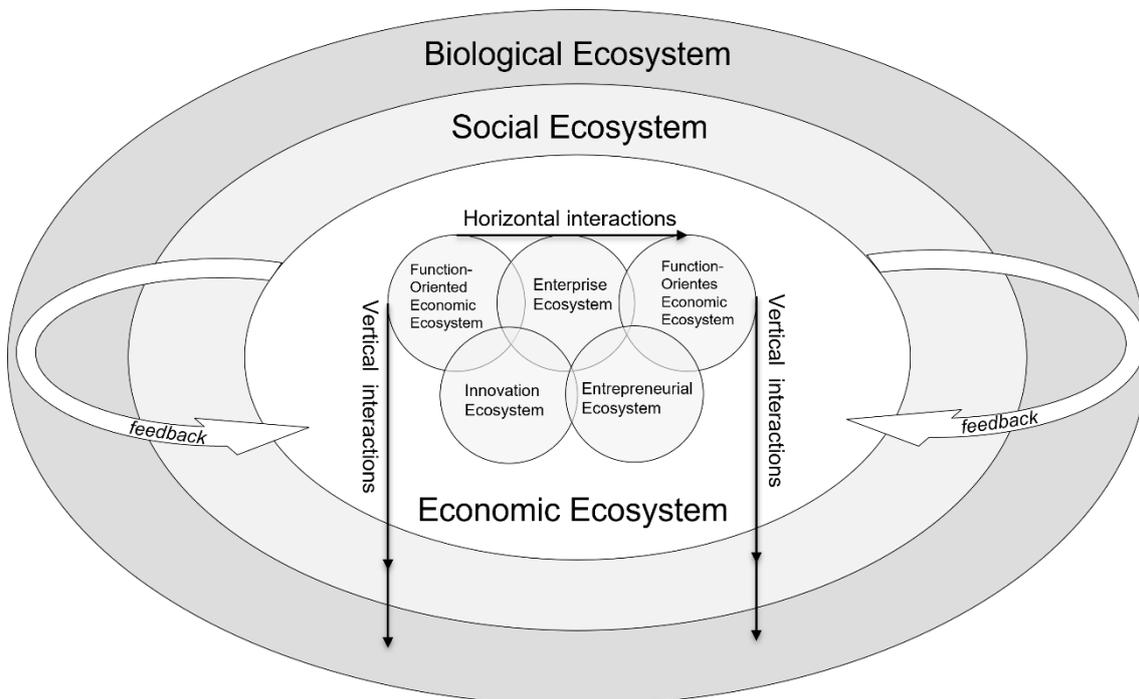

**Figure 1 |** Panarchic Integration Framework of Economic, Social, and Biological Ecosystems. *Source: Author's own elaboration.*
The diagram illustrates the hierarchical embedding of economic ecosystems within broader social and biological ecosystems, reflecting a panarchic structure. Vertical interactions represent cross-level dependencies, while horizontal interactions capture dynamic feedbacks across economic sub-ecosystems (enterprise, innovation, entrepreneurial, and function-oriented ecosystems). The framework emphasizes the systemic interdependence underlying economic dynamics as embedded real subsystems within adaptive complex systems.

***Emergent failures and structural externalities***

Traditional economic theory frames externalities as market failures—anomalies external to the price mechanism (*155–157*). Yet from a complexity perspective—and when viewed through a panarchic lens—these are not anomalies but emergent phenomena arising from dynamic interactions within and across nested systems. Environmental degradation, for instance, is not merely an unpriced cost but a cumulative effect of non-beneficial systemic interactions (*158–160*). Likewise, market concentration and inequality are not deviations from ideal structures, but structural symptoms of competitive dynamics that erode diversity and disrupt systemic equilibrium (*18, 20, 161*).

*Unmasking the Illusion of Mutualism*

One reason such failures persist is the enduring belief that market exchange inherently produces mutual benefit. Early classical economists framed voluntary exchange as a cooperative mechanism aligned with the public good (*1, 162, 163*), and this assumption became foundational in mainstream economic thought. Friedman (*3*) later reinforced this logic, claiming that unrestricted individual gain would naturally translate into social welfare. A similar notion emerged in evolutionary theory through the concept of Evolutionarily Stable Strategies (ESS) (*164, 165*), which posits that under certain conditions, self-interested behavior can generate system-level equilibrium.

However, such conditions are rarely met in real-world economies. Mainstream capitalism, increasingly equating competition with mutual benefit (*3, 5*), institutionalizes structurally non-beneficial interactions—from financialization and rent extraction (*164*) to platform monopolies (*20*) and systemic power concentration (*14*). What is framed as mutualism, in structural terms, often manifests as predation, parasitism, or destructive competition.

*A formula for ecosystemic equilibrium*

As presented earlier, empirical studies show that beneficial interactions account for the majority of relationships in resilient biological systems—often exceeding 70% (118, 120, 121, 126, 127). In contrast, the dominance of non-beneficial interactions—particularly sustained competition—leads to reduced diversity, monopolization, and the collapse of ecosystemic equilibrium (121, 130–133). These patterns are mirrored in economic systems, where non-beneficial behaviors increasingly shape market dynamics (18, 20, 166, 167).

We propose a theoretical model to assess the ecosystemic equilibrium of an economic system:

$$E = \frac{\alpha_B \cdot R_B \cdot D}{\alpha_N \cdot R_N}$$

Where:
- $R_B$: Sum of beneficial relationships.
- $R_N$: Sum of non-beneficial relationships.
- $\alpha_B$: Intensity coefficient of beneficial relationships.
- $\alpha_N$: Intensity coefficient of non-beneficial relationships.
- $D$: Diversity index of the ecosystem.

**E > 1** indicates systemic equilibrium; **E < 1** indicates fragility. This framework builds on ecological models where stability emerges from the predominance of beneficial interactions and diversity (*72, 73, 114–117, 168*). It enables diagnostic use in policy, market assessment, and organizational strategy.

Parameter estimation can leverage network mapping, agent-based modeling, and survey-based indices of cooperation or conflict (*15, 70*). Future work may simulate thresholds under different RB/RN dynamics using computational ecosystems.

This model is not intended to predict system behavior but to describe structural conditions that favor ecosystemic equilibrium. It offers a diagnostic tool for assessing the balance of beneficial and non-beneficial interactions within a given system, rather than a deterministic equation of outcomes.

**Conclusion: from competitive collapse to ecosystemic equilubrium**

This work extends ecological interaction typologies into the realm of economics, demonstrating that economic ecosystems are not metaphorical constructs but real complex adaptive systems—structurally isomorphic with biological and social ecosystems. This structural equivalence enables a reconceptualization of systemic dynamics often overlooked in conventional economic models, especially those involving feedback, resilience, and emergence.

As shown throughout this study, economic systems function as nested subsystems within broader social and biological ecosystems. Within this panarchic architecture, and by theoretical extension, competition must be reclassified—not as a beneficial force, but as a non-beneficial interaction due to its long-term effects. Empirical evidence confirms that competitive dynamics lead to exclusion, concentration, and depletion—undermining ecosystemic equilibrium, social cohesion, and environmental resilience. While competition may persist as a natural interaction, its role in economic, business, and policy frameworks must be constrained and subordinated to systemic resilience.

This is not a normative stance, but a systemic one: resilient ecosystems are dominated by beneficial interactions such as mutualism, cooperation, and facilitation. These interactions enhance adaptability and robustness, while competitive dominance correlates with fragility and collapse.

Reframing economies as ecologically embedded systems clarifies the origins of many so-called "market failures." From a complexity perspective, these are not anomalies but emergent consequences of structurally unbalanced interaction patterns propagating across nested ecosystems.

To support this reconceptualization, we have proposed a theoretical model of ecosystemic equilibrium. By assessing the ratio and intensity of beneficial versus non-beneficial interactions, along with system diversity, this framework enables structural diagnostics of resilience and fragility. While preliminary, the model offers a foundation for future empirical research—ranging from agent-based simulations to network mapping and firm-level assessments.

The implications extend beyond theory. Institutional design and public policy must shift away from fostering competition for short-term performance under equilibrium assumptions, and instead promote conditions that enable beneficial mechanisms, distributed agency, and long-term systemic resilience. Where competition persists, it should be strictly bounded and subordinated to the preservation of ecosystemic equilibrium.

Ultimately, this framework challenges the foundations of competition-centered economic thought. It redefines economic systems as ecologically embedded, socially interdependent, and structurally adaptive ecosystems—where sustainability is not an externality, but a function of design. In doing so, it opens a pathway toward economic models capable of supporting resilience, equity, and ecosystemic equilibrium.

Table 1. Methodological Composition of the Reviewed Literature

| Disciplinary Domain | Total Works | Foundational Works | Recent Article. | Landmark Works 2001-2018 | Selection Criteria |
|---|---|---|---|---|---|
| *Biology and Ecology* | 46 | 5 | 18 | 23 | Influence in the Field, Theoretical or Empirical Support, Citations. |
| *Economic and Business Science* | 75 | 16 | 36 | 23 | Influence in the Field, Theoretical or Empirical Support, Citations. |
| Complexity Sciences | 37 | 5 | 19 | 13 | Influence in the Field, Theoretical or Empirical Support, Citations. |
| *Philosophy of Science* | 10 | 5 | 3 | 2 | Influence in the Field, Theoretical or Empirical Support, Citations. |

Source: Own elaboration

Table 2. Typology of Ecosystem Interactions and Their Long-Term Effects

| Type of Interaction and Long-Term Effect | Type of Interaction | Interaction Sign *(Effect-Based)* | Description of the Effect on the Participants |
|---|---|---|---|
| **Beneficial** | Mutualism | +/+ | All participating organisms benefit from the interaction, and none are harmed. It occurs between different species and involves a degree of interdependence. |
| | Cooperation | +/+ | All participating organisms benefit from the interaction, and none are harmed. It can occur within or between species and does not imply interdependence. |
| | Commensalism | +/0 | One or more organisms benefit from the interaction without negatively affecting any other. |
| | Facilitation | +/+/0 | One or more organisms modify the environment in ways that benefit others. |
| **Non-Beneficial** | Predation | +/- | One or more organisms benefit from the interaction while the others are harmed. |
| | Parasitism | +/- | One or more organisms benefit at the expense of others, which are harmed by the interaction. |
| | Amensalism | 0/- | One or more organisms are unaffected by the interaction, while others are harmed. |
| | Competition | -/- | All participants are negatively affected as they compete for limited resources. |

Source: Adapted by the authors from ecological interaction mechanisms (118–122).

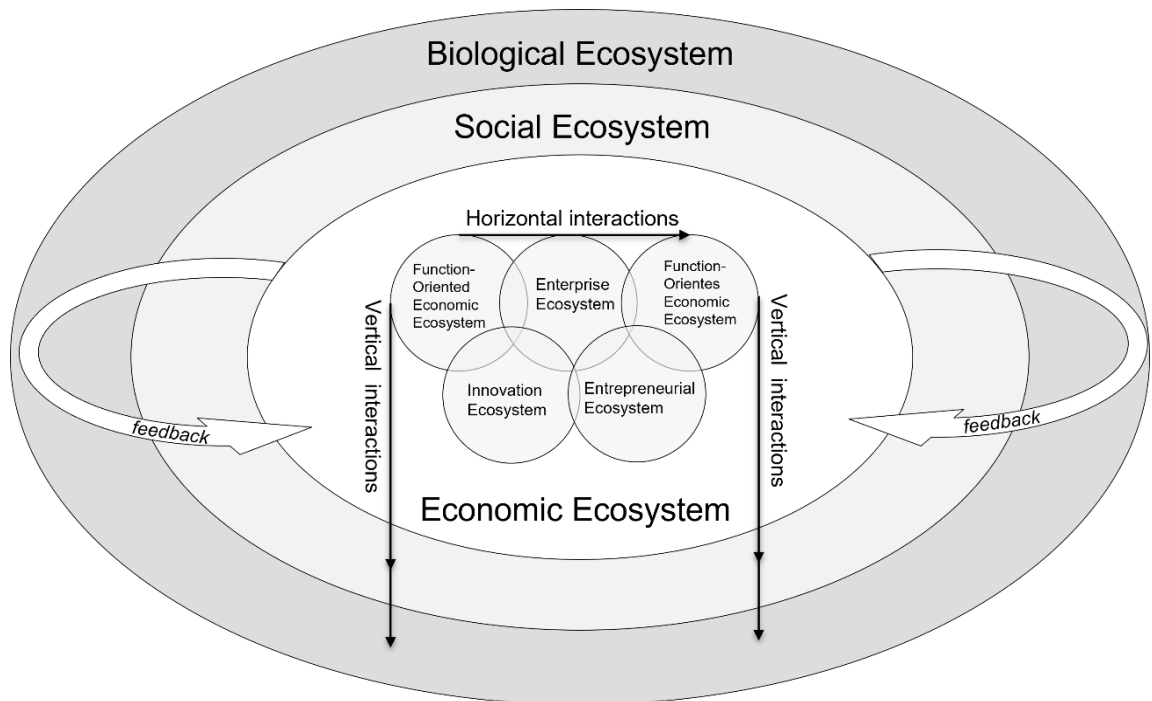

**Figure 1** | Panarchic Integration Framework of Economic, Social, and Biological Ecosystems.
*Source: Author's own elaboration.*
The diagram illustrates the hierarchical embedding of economic ecosystems within broader social and biological ecosystems, reflecting a panarchic structure. Vertical interactions represent cross-level dependencies, while horizontal interactions capture dynamic feedbacks across economic sub-ecosystems (enterprise, innovation, entrepreneurial, and function-oriented ecosystems). The framework emphasizes the systemic interdependence underlying economic dynamics as embedded real subsystems within adaptive complex systems.